# Energy and exergy analysis of solar stills with micro/nano particles: A comprehensive study.

S.W. Sharshir[1,2,3], Guilong Peng[1,2], A.H. Elsheikh[4], Talaat A. Talaat[5], Mohamed A. Eltawil[6], A. E. Kabeel[7], Nuo Yang[1,2]*

[1] State Key Laboratory of Coal Combustion, Huazhong University of Science and Technology, Wuhan 430074, China

[2] Nano Interface Center for Energy (NICE), School of Energy and Power Engineering, Huazhong University of Science and Technology, Wuhan 430074, China

[3] Mechanical Engineering Department, Faculty of Engineering, Kafrelsheikh University, Kafrelsheikh 33516, Egypt

[4] Production Engineering and Mechanical Design Department, Faculty of Engineering, Tanta University, Tanta, Egypt.

[5] Physics and Mathematical Engineering Department, Faculty of Electronic Engineering, Menoufiya University, Egypt

[6]Agricultural Engineering Department, Faculty of Agriculture, Kafrelsheikh University 33516, Egypt

[7] Mechanical Power Engineering Department, Faculty of Engineering, Tanta University, Tanta, Egypt.

*Corresponding authors: Nuo Yang. (nuo@hust.edu.cn)

# Abstract

In this paper, a comparative study between modified solar stills (with graphite or copper oxide micro/nano particles) and classical solar still is carried out, based on the productivity and the thermal performance. Exergy destructions in various components of the solar stills have been calculated, analyzed and discussed. Evaporation is faster and the exergy of evaporation is higher at the modified solar stills than that of the classical one. Furthermore, the energy and exergy efficiencies of the modified stills are enhanced compared with the classical one. A brief discussion regarding the effect of different parameters on solar stills efficiency is also presented. The daytime energy efficiency of graphite/water and copper oxide/water mixtures are 41.18% and 38.61%, respectively, but for the classical still is only 29.17%. Moreover, the daytime exergy efficiencies of graphite, copper oxide nanofluid based stills and classical still are 4.32%, 3.78% and 2.63%, respectively.

## 1. Introduction

Only 1% of obtainable water resources on the earth is drinkable, 2% is frozen in polar glaciers and the remnant of them are brackish and saline (97%) [1]. Solar stills (SSs) can be used as one of the most suitable solar desalination units for low-cost, low-capacity, simple-to-operate and self-reliant water supply systems [2]. The achievement of SSs, as a type of renewable energy conversion systems, is evaluated according to exergy and energy dissection which depending only on the accounting of energies entering and exiting the system. Due to low yield of the SSs, many investigators introduced several scrupulous designs to enhance the output by enhance the evaporation rate of water. Researchers suggested different designs to improve the SSs' performance, such as double slope [3], stepped SS [4, 5], SS with phase change materials (PCMs) [6-9] wick SS [12-10], hybrid SS [13-15], stepped SS with continual water circulation [3], wind speed [16], SS with storage materials [17, 18], SS with nanoparticles [19], SS with PV/T [20-22] , and SS with vertical ripple surface [23].

The major parameters influenced the yield of SS are the heat transfer mechanism and operating temperature. The coefficient of heat transfer can be improved by enhancing the thermal properties of the base water. Suspending nanoparticles into the base fluid is a very simple technique which enhances the thermal attribute and hence the productivity is also enhanced. Nanofluid is formed by suspending nano-sized particles in a base fluid. The pioneer work was performed by [24, 25] on the improvement of fluids thermal conductivity utilizing nanoparticles. Several investigators studied the influences of utilizing various kinds of nanofluid on the output (productivity) of SS. The yield of the SS with $Al_2O_3$, $SnO_2$ and $ZnO$ nanofluids was improved by 29.95%, 18.63% and 12.67% higher than that without nanofluids, respectively [26]. The influences of utilizing $Al_2O_3$ and $CuO_2$ nanoparticles and a condenser integrated with a SS under the Egyptian conditions were investigated by [27, 28]. Results illustrated that an enhancement of 53.2% in the total diurnal yield with vacuum inside the SS and an enhance of 116% when utilizing the $Al_2O_3$ nanoparticles with vacuum as well as $CuO_2$ nanoparticles improved the yield by 133.6% and 93.8% with and without with vacuum. The productivity of the SS with $Al_2O_3$ nanofluid was increased by 12.2% and 8.4% at 35 kg and 80 kg bottom fluid mass, respectively, with 0.12% wt of $Al_2O_3$ nanoparticles [29]. Sharshir et al. [30] experimentally investigated the effect of graphite micro/nano -flakes (GMF) and copper oxide (CuO) particles with glass cover cooling on the SS performance. The GMF (CuO) with various weight concentrations, brine depths, and glass cover cooling flow rates were examined. The optimal daily efficiencies of the SSs with GMF (CuO) are 49% (46%), respectively, compared with the classical value as 30%.

The quantity of energy in SSs is studied according to the $1^{st}$ law of thermodynamics. While the quality of energy is represented by exergy analysis according to the $2^{nd}$ law of thermodynamics. Exergy is a strong instrument to identify the causes, locations and magnitude of the system inefficiencies. In addition, it provides a precision measure how the SS approaches to the ideal [31, 32]. It was found that for all renewable conversion energy systems, the evaluation according to exergy analysis was establish to be less significant than that of the energy analysis [32]. Dwivedi and Tiwari [33] presented thermal analysis for active SS on the basis of energy analysis. It has been observed that the still of double slope active type under natural modes gives 51% higher productivity compared to the still of double slope of passive type. Vaithilingam and Esakkimuthu [34] conducted experiments with different water depths. The influences of water depth on energy and exergy efficiencies and exergy destruction of various components of the SS were studied. Ranjan et al. [35] carried out an exergy and energy evaluation on SSs. It was spotted that the exergy efficiency is smaller than the energy efficiency.

They found that the exergy destruction rate in the still components is much dependent on the rate of incident insolation with evaluated values equal to 9.7, 62.5 and 386 W/m$^2$ for the glazier, the water and the bottom trough, respectively. Shanmugan et al. [36] studied, experimentally, the performance of SS and evaluated the instantaneous exergy and energy efficiencies during summer and winter. Tiwari et al. [37] compared between passive and active SS based on the hourly yield. The influence of the number of collectors and the brine depth on exergy and thermal efficiency of the passive and active SS is obtained. The results showed that: as the number of collectors and brine depth are increased, the thermal efficiency decreases. In our previous studies [6, 30], the effect of using different nanofluids with different concentrations on the solar still productivity has been investigated. However, for our best knowledge, the thermal performance such as energy efficiency, exergy efficiency and exergy dissipation on solar stills with nanofluids has not been investigated.

The present study aims to investigate the influences of nanoparticles on the energy efficiency, exergy efficiency and exergy dissipation on SS components (basin, glazier and water) with and without nanofluids as well as the convective, evaporative and radiative heat transfer coefficients on the performance of the SS with and without nanofluids. Enhancing the exergy of evaporation and productivity of modified SS (with graphite and copper oxide (CuO) nanofluids) compared with classical SS is investigated. Moreover, augmenting the diurnal exergy efficiencies of the SS components, i.e., the basin liner, the glazier cover and the saline water by utilizing nanofluids is also presented.

## 2. Experimental setup and devices

Three similar SSs were designed and fabricated to assess the total performance of the units. The solar desalination setup consists of a classical solar still (CSS), a modified solar still (MSS) with graphite nanofluid, and an MSS with CuO nanofluid. The stills are made of iron sheets. To enhance the absorptivity of the SS and hence improve the rate of evaporation, a black paint is used to coat all inner basin surfaces. To eliminate heat losses as possible, all external and bottom surfaces are insulated by fiberglass. The basin was covered with glass sheet inclined at nearly 30° horizontally, which is the latitude of Wuhan, Hubei, China. More details regarding the experimental setup including a sketch, a photo, and measuring instruments specifications can be found in [6, 30].

Experiments were conducted at the school of energy and power engineering, Huazhong University of Science and Technology, Wuhan, China and carried out from 9 am to 5 pm during the period of October and November 2015. During the experiments, total solar intensity, air velocity, the glass and brine water temperatures, ambient temperature, and the distilled water amount are measured. All measured parameters are measured every 1 h.

Measuring instruments specifications, depending on the commercial types used through the study, as well as their accuracy, range and computed experimental errors are tabulated in Table 1. The uncertainties in the obtained experimental results are calculated according to the model proposed by [38]. The uncertainty limits of the temperature measurement were about 0.05 °C which calculated according to the following equation considering the freezing and boiling temperatures for water are 0 °C and 100 °C [39]:

$$S_T = \left[ \left( \frac{\partial T_a}{\partial T_{b,m}} S_{T_{b,m}} \right)^2 + \left( \frac{\partial T_a}{\partial T_{f,m}} S_{T_{f,m}} \right)^2 + \left( \frac{\partial T_a}{\partial T_m} S_{T_m} \right)^2 \right]^{1/2} \qquad (1)$$

where $T_a$, $T_{b,m}$, $T_{f,m}$, and $T_m$ are the actual, measured boiling, measured freezing and measured temperatures, respectively. And $S_{T_{b,m}}$, $S_{T_{f,m}}$, and $S_{T_m}$ are the uncertainties in $T_{b,m}$, $T_{f,m}$, and $T_m$, respectively, and they have the same values for all thermocouples used in experiments.

Table 1 Measuring instruments specifications.

| Measured parameter | Instrument | Range | Accuracy | % Error |
|---|---|---|---|---|
| Temperature | Calibrated copper constantan type thermocouples connected to a digital temperature indicator (model TES-1310) | -50 to 280 °C | ± 1 °C | 0.5 |
| Solar intensity | solar meter (model TES-1333) | 0-2000 W/m² | ± 10 W/m² | 0.25 |
| Air velocity | Vane type digital anemometer (Model Benetech GM816) | 0-30 m/s | ± 1 m/s | 5 |
| Productivity | A graduated cylinder | 1.5 l | ± 0.002 l | 10 |

## 3. Heat transfer analysis in solar stills

Generally, the heat transfer process in a SS is considered as either internal or external heat transfer processes depending on energy transferred in or out the SS enclosure [40-42]. The internal heat transfer takes place inside the SS within its enclosed space and is responsible for the transport of pure water in vapor form leaving impurities behind in the basin itself. Internal heat transfer is mainly depending on the nanofluid thermo-physical properties. While, the external heat transfer occurs across the bounded space of the solar still, precisely, through its condensing cover and is responsible for the condensation of pure vapor as distillate. Both internal and external heat transfer processes, as well as, the thermo-physical properties of the nanofluids, are briefly described in the subsequent sections.

**3.1 Water-nanoparticles mixture thermo-physical properties**

Nanofluids have many superior properties compared to its base liquid such as high thermal conductivity, high insolation absorptivity, which may improve the productivity of the SSs [30]. Most physical properties of nanofluids can be described as functions of the base fluid and the nanoparticles properties. **The equations of the main properties of nanofluids are given in** [43-45] as:

nanofluids density [43]:

$$\rho_{nf} = \left(\frac{\varphi}{100}\right)\rho_p + \left(1 - \frac{\varphi}{100}\right)\rho_{sw} \tag{1}$$

wherever $\varphi$ is the weight percentage and is given by:

$$\varphi = \left(\frac{m_p}{m_p + m_{sw}}\right) \times 100 \tag{2}$$

nanofluids specific heat [43]

$$C_{nf} = \frac{(\varphi/100)\rho_p C_p + (1-(\varphi/100))\rho_{sw} C_{sw}}{\rho_{nf}} \tag{3}$$

nanofluids thermal conductivity [45]

$$K_{nf} = K_{sw} \frac{K_p + (n-1)K_{sw} - (n-1)\varphi(K_{sw} - K_p)}{K_p + (n-1)K_{sw} + \varphi(K_{sw} - K_p)} \quad (4)$$

Inside the SS enclosure, the convection and radiation modes of heat transfer are occurred simultaneously. The heat energy is lost to the ambient from the outer surface of the glazier cover by radiation and convection heat transfer processes. All related heat transfer equations could be found in our previous work [32].

### 3.2 Energy efficiency of solar still

Energy efficiency is considered as one of the most important criteria used to assess the performance of SSs [32]

The total efficiency, $\eta$, is obtained by the summation of the hourly yield $m_{ew}$ multiplied by the evaporation latent heat $h_{fg}$ divided by the diurnal average insolation $I(t)$ over the whole area $A_s$ of the device

$$\eta = \frac{\sum m_{ew} \times h_{fg}}{\sum I(t) \times A_s \times 3600} \quad (5)$$

The balance equations of exergy for the three main components (basin-liner, saline water and glazier cover) of the SS are given in our previous work [32]. Assuming that the heat capacity of the components materials is negligible, and hence the accumulation of exergy could be neglected. The quality of energy is represented by exergy analysis based on the second law of thermodynamics. The exergy analysis is a powerful tool to identify the causes, locations and magnitude of the system inefficiencies. In addition, it provides a precision measure how the SS approaches to the ideal. The exergy efficiency of SS ($\eta_{EX}$) is defined as the ratio between the exergy output associated with the distillate water to the exergy input of radiation [32].

$$\eta_{EX} = \frac{\text{Exergy output of solar still}}{\text{Exergy input of solar still}} = \frac{E_{x_{evap}}}{E_{x_{input}}} \tag{6}$$

In a SS, exergy output is a result of evaporation and subsequently the condensation of saline water. In practice, some of the evaporated water after condensation on the cover, falls back into the basin, hence the evaluated exergy output from the experimental results would be less than the theoretical one. The hourly exergy output of a SS can be defined as [32]:

$$E_{x_{output}} = E_{x_{evap}} = \frac{\dot{m}_{ew} \times h_{fg}}{(3600 s.h^{-1})} \times \left(1 - \frac{T_a}{T_w}\right) \tag{7}$$

wherever, $\dot{m}_{ew}$ is hourly yield of SS (kg/h), $h_{fg}$ is the latent heat (J/kg), $T_a$ is the surrounding air temperature (°C) and $T_w$ is the basin water temperature (°C).

The exergy input to SS through radiation $E_{x_{sun}}$ can be expressed in terms of insolation as follows [32]:

$$E_{x_{sun}} = A_s \times I(t)_s \left[1 - \frac{4}{3} \times \left(\frac{T_a}{T_s}\right) + \frac{1}{3} \times \left(\frac{T_a}{T_s}\right)^4 \right] \tag{8}$$

Wherever, $A_s$ is the area of the basin in (m²), $I(t)_s$ is the insolation on the inclined glazier surface of the SS (W/m²) and $T_s$ is the sun temperature (6000 K).

## 4. Results and discussion

### 4.1 Materials and Nanofluids Characterization

The morphology of the graphite micro/nano -flakes and the copper oxide particles were measured by scanning electron microscope (Helios Nanolab G3 CX) (See Fig. 1). It is obvious that the graphite particles have large lateral size and small thickness, hence it has a flake shape. The graphite micro/nano -flakes particles are 10000 meshes, they are purchased from the local market where its price is around $30/kg as well as the CuO price is around $60/kg. The average lateral size of the flakes is about 1.3 μm, the thickness is around 100 nm as shown in Fig.1-a. The morphology of CuO particles is cuboid, where their size is around 1 μm. Unlike graphite particles, the copper oxide particles aggregated tightly, the size of the aggregated particles is around 1 μm as shown in Fig. 1-b as well as the specifications of graphite and CuO particles are shown in Table 2.

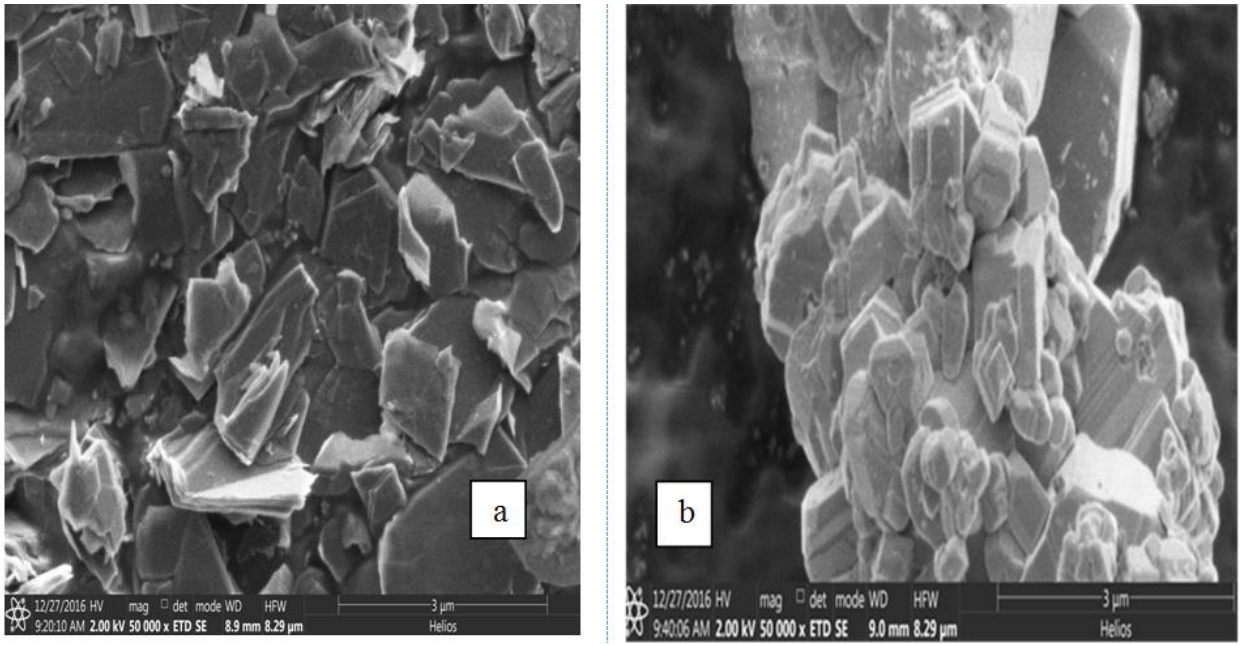

Fig.1 SEM images of a) graphite micro/nano-flakes and b) copper oxide micro particles.

Table 2 Specifications of nanoparticles

| Properties | Graphite | Copper oxide |
| --- | --- | --- |
| Thermal Conductivity, W/ (m.K) | 129 | 76 |
| Density, g/cm$^3$ | 1.2 | 6.4 |
| Particle size average, µm | 1.2~1.3 | ~1 |

Compared with the absorbing plate surface, the absorptivity of the 3D absorbing materials like porous material and nanofluid is high due to the multiple scattering and absorption of the light. The absorption characteristics of the materials used in this study were measured by UV-Vis (LabRAM HR800) at (400 to 1100 nm wavelength) as shown in Fig. 2 compared with the absorption of the black basin 94%, the absorption of the uniform graphite micro/nano -flakes is around 90%, lower than that of the black paint by 4%, whereas the absorption of copper oxide particles is around 91.5%, which is lower than that of the black paint by 2.5% and higher than that of the graphite micro/nano-flakes by 1.5%. This absorption is given until 850 nm wavelength after that, from 850 to 1200 nm the copper oxide particles decrease to about 70% as shown in Fig. 2-a. Furthermore, Fig. 2-b. illustrates the absorption of pure water, 0.5% graphite nanofluids and 0.5 copper oxide nanofluids. From Fig.2-b the absorption of the 0.5% uniform graphite nanofluid is around 99.5%, which is higher than that of the black paint (~94%) by 5.5% and higher than that of graphite micro/nano -flakes by about 9.5%. This is because of the fact that the graphite nanofluid has a 3D absorbing structure, which is good for trapping the light to be absorbed for many times on the graphite flakes surface.

On the other hand, the absorption of 0.5% copper oxide nanofluid is around 96%, which is higher than that of the black paint (~94%) by 2% and higher than that of copper oxide particles by about 4.5%. This is because of the fact that the copper oxide nanofluid has a 3D absorbing structure, which is good for trapping the light to be absorbed for many times on the graphite flakes surface.

The nanoparticles have a small size with a very large surface area, which result in a significant increase in the absorption of the solar energy. The thermal conductivity and absorption of the nanofluids is significantly high compared to the base fluid due to the presence of nanoparticles [6, 43]. Nanofluids, in general, have high density and high convective heat transfer coefficient (HTC) with a low specific heat of nanoparticles which result in increasing thermal conductivity of nanofluid. Increasing the nanofluid thermal conductivity leads to increase the HTC which in turn increases the amount of evaporation. As a result, more productivity can be obtained.

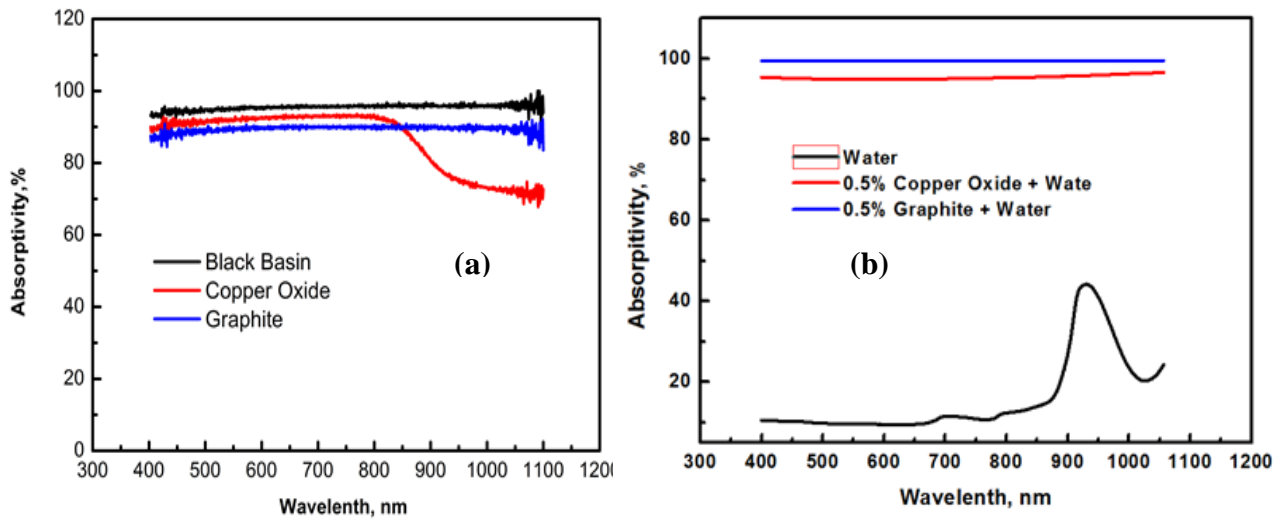

Fig. 2. (a) The absorption of the black paint graphite micro/nano -flakes [6] and copper oxide

particles. (b) The absorption of the water, and 0.5% uniform graphite [6] and 0.5% copper oxide with water.

**4.2 Effect of meteorological parameters on the solar stills performance**

Effects of CuO and graphite nanofluids on the performance of still at water depth equal to 0.5cm and concentration of 1% will be discussed here. The hourly insolation, air velocity and air temperature for one day of experiments, and components temperature are illustrated in Table 3. Many points could be figured out from Table 3. and will be stated in the subsequent illustrations. All temperatures of brine water, plate and glazier are raised gradually with the raise of insolation and reach a peak value at approximately 13 p.m., then the insolation and ingredients temperatures are depressed. In addition, the SS hourly productivity is also proportionate to the insolation. The brine and basin plate temperatures of the still with graphite nanofluid are more than that of the CSS by 1.5 – 4.5ºC and 0.9 – 4 ºC, respectively. The brine and plate basin temperatures of the CuO nanofluid still are more than that of the CSS by 1.6 – 4.1ºC and 1.2– 4 ºC, respectively. The temperature of the inner and outer surface of the glazier cover for CuO nanofluid still are more than that of the CSS by 0.5–3ºC and 0.5 – 2 ºC, respectively. The temperature of the inner and outer surface of the glazier cover for graphite nanofluid still are more than that of the CSS by 0.2–3.5ºC and 0.5 – 2 ºC, respectively. Hence, the evaporation and production rates are better in modified stills (with nanofluids) than that of CSS (without nanofluids).

Table 3. Hourly experiments measurements.

| | Meteorological parameters | | | CSS | | | | SS with CuO | | | | SS with Graphite | | | |
|---|---|---|---|---|---|---|---|---|---|---|---|---|---|---|---|
| Time | I(t) (W/m$^2$) | Ta (K) | Va m/sec | $T_b$ | $T_w$ | $T_{go}$ | $T_{gi}$ | Tb1 | Tw1 | Tgo1 | Tgi1 | Tb2 | Tw2 | Tgo2 | Tgi2 |
| 9 | 440 | 299 | 2.1 | 316.7 | 313.5 | 309 | 310.5 | 319 | 315.6 | 310 | 311 | 318 | 315.35 | 311 | 310.3 |
| 10 | 650 | 301 | 2 | 332.4 | 328.5 | 321 | 323.5 | 336.4 | 332.6 | 323 | 325.5 | 336.4 | 332.75 | 323 | 325.2 |
| 11 | 775 | 302 | 1.9 | 342.9 | 338.5 | 330 | 332.5 | 346.5 | 341.6 | 331 | 333.5 | 345.5 | 340.85 | 331.8 | 332.5 |
| 12 | 900 | 303 | 2.2 | 347.2 | 342.5 | 332 | 336.5 | 351 | 345.6 | 334 | 337.5 | 351 | 345.85 | 334 | 337.6 |
| 13 | 880 | 304 | 2.6 | 349.1 | 344.5 | 334 | 337.5 | 352.5 | 347.6 | 335 | 339.5 | 352.5 | 347.85 | 335.5 | 339.4 |
| 14 | 820 | 303 | 1.7 | 343.7 | 339.5 | 330.5 | 332.5 | 346 | 341.6 | 331 | 333.5 | 345 | 341 | 331.5 | 332.5 |
| 15 | 695 | 303 | 1 | 332 | 329.5 | 321 | 324.5 | 335.5 | 333.1 | 322 | 327.5 | 336 | 334 | 323 | 328 |
| 16 | 440 | 302 | 0.7 | 324.3 | 322 | 315 | 318 | 325.5 | 323.6 | 315.5 | 319 | 326 | 324.45 | 316.5 | 319.5 |
| 17 | 252 | 301 | 1.1 | 317.6 | 315.5 | 311 | 313.5 | 319.5 | 317.6 | 311.5 | 314.5 | 318.5 | 317.1 | 311.5 | 314 |

**4.3 Hourly convective, evaporative and radiative heat transfer coefficients**

The hourly convective, evaporative and radiative heat transfer coefficients (HTCs) rates of CSS and MSSs with graphite and copper oxide particles at a weight concentration of 1% and water depth 0.5cm are clarified in Fig. 3. The convective and evaporative HTC have significance values of MSSs with graphite and CuO particles than CSS. Convective HTC of CSS varies from 1.2 to 2.6 W/m² K, MSS with CuO varies from 1.2 to 2.6 W/ m² K and MSS with graphite varies from 1.2 to 2.65 W/ m² K respectively as illustrated in Fig. 3-a, this due to the higher temperature in MSSs compared with CSS. Furthermore, the evaporative HTC values of MSSs changes from 8 to 58 W/ m² K for graphite nanoparticles and from 8 to 56 W/ m² K for CuO microparticles as well as 8 to 47 W/ m² K for CSS respectively as illustrated in Fig. 3-b, also the higher evaporative HTC in MSSs also due to due to the higher temperature in MSSs compared with CSS.

Moreover, from Fig. 3-c, d shows the hourly rate of radiative HTC between the water and glazier as well as between glazier and the atmosphere. The radiative HTC between the water and glazier in MSSs a little higher than CSS due to the lower temperature difference between the water and the glazier.

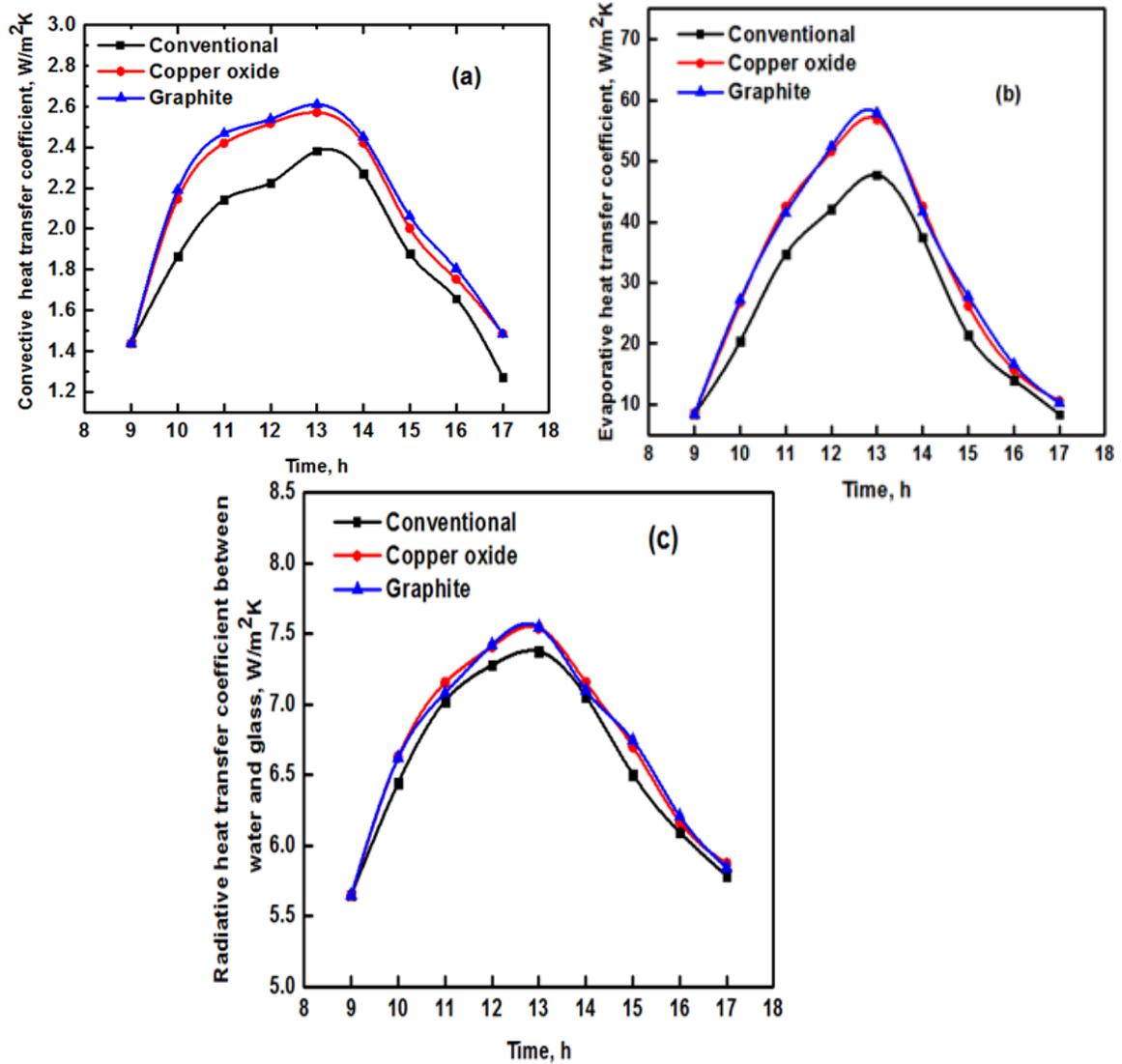

Fig.3 Hourly heat transfer coefficients (HTC) rate of CSS and MSSs still (a) Convective, (b) Evaporative and (c) Radiative between water and glazier.

**4.4 Exergy of evaporative and freshwater productivity**

Evaporative heat flux is required to evaporate water from the water surface in the basin liner area. The evaporation exergy of the three SSs under study are shown in Fig. 4-a, this figure reveals that the exergy evaporated is higher at 13 pm for both stills (with and without nanofluids). These values are found to be 14.33, 13.70, and 10.23, W for the nanofluids (graphite and copper oxide) and CSS, respectively. For the still with nanoparticles (graphite and copper oxide) the water temperature is higher. As a result, evaporation is faster and the exergy of evaporation is higher at stills with graphite and CuO nanofluids. Hence, the productivity of the nanofluids stills is higher than that of the CSS.

The variations of hourly freshwater productivity for MSSs and CSS are presented in Fig.4-b. It can be observed from Fig. 4-b that the productivity trend is similar to that of the exergy of evaporation as observed from Fig.4-a. The main reason of the productivity increase in the modified stills compared with the CSS is the significant increase in the evaporation rate, as the heat transfer rate is increased by utilizing nanofluids. The diurnal productivity of MSSs the graphite and CuO nanofluid are enhanced by 41.18% and 32.35%, respectively, compared with CSS.

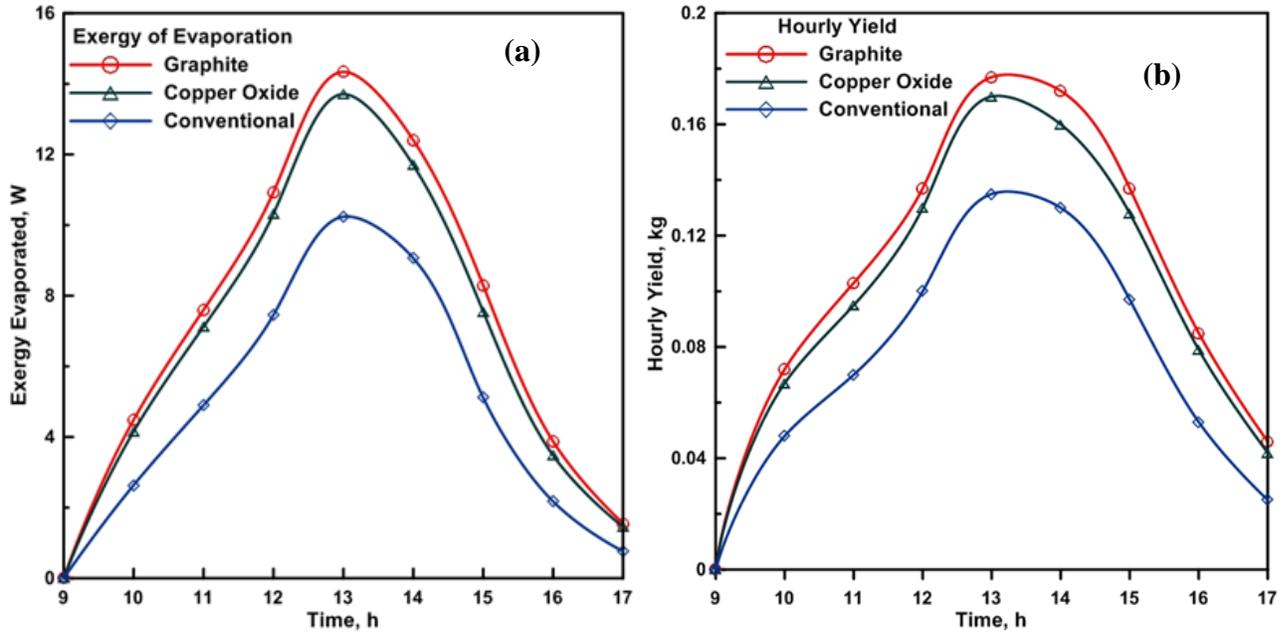

Fig.4. (a) Hourly variation of exergy evaporated with and without nanofluids. (b) Hourly freshwater productivity with and without nanofluids.

**4.5 Exergy dissipation and exergy efficiency in still components**

The quantity and location of exergy destruction could be identified through exergy analysis. Therefore, the exergy efficiency would be enhanced by applying suitable measures and reduce the exergy destruction. The rate of instantaneous exergy destruction was calculated for different components of the SSs with and without nanofluid such as basin liner, saline water, and glazing cover.

Variation of exergy destruction in basin liner based on hourly intervals was shown in Fig.5-a. the highest daily exergy destruction of (12384 kJ/m$^2$ day) in basin liner with and without nanofluids was almost the same at 0.5 cm water depth and 1% concentration of nanofluids. Moreover, as shown in that figure the exergy destruction of basin liner of SS with nanofluids was a bit little larger than that of classical still due to the lowest temperature difference between basin liner and water ($T_b - T_w$) in MSSs compared with CSS. The exergy destruction values in the present study are in agreement with Zoori et al. [46]. The highest exergy destruction was observed in the basin liner. This may refer to lower difference in temperature between basin liner and water ($T_b - T_w$).

Fig. 5-a shows the exergy dissipation that occurs in the glazing cover. It was found that, the glazing cover exergy dissipation in modified SSs was higher than that of CSS. This may be due to higher difference in temperature between the glazing cover and ambient in the MSSs compared to the CSS. The maximum daily exergy destruction of the glazier cover occurs in the still with graphite and CuO nanofluids and CSS are 1492, 1419 and 911 kJ/m$^2$ day,

The lowest exergy destruction was obtained in the saline water as shown in Fig. 5-a. It was noticed that the exergy destruction in the saline water was decreased by increasing the temperature difference between saline water surface and the internal surface of the glass ($T_w - T_{gi}$). This increase in the temperature difference increases the exergy evaporated ($E_{xtw\_g}$) from the water surface and reduces the exergy dissipation of saline water. This in agreement with Vaithilingam [34]. Hence, the exergy dissipation of saline water with nanofluid was lower than that of the still without nanofluid due to the higher difference in temperature in the former compared with the latter case. Therefore, the evaporation rate in the modified SSs was larger than the CSS.

The high difference in temperature between basin liner and water leads to increase the exergy associated with water ($E_{xw}$) and decrease the exergy dissipation in the basin liner. The maximum diurnal exergy destruction of the saline water occurs in the nanofluids (graphite and copper oxide) SSs and CSS are 473, 787, and 1075 kJ/m² day, respectively.

Furthermore, Fig.5-a shows the exergy losses through insulation $E_{X_{ins}}$ for the MSSs and CSS. The daily insulation exergy losses in the MSSs are 1693.872 kJ/m² day and 1664.48 kJ/m² day for CuO and graphite, respectively, which is higher than that of the classical one (1447.128 kJ/m² day) by 17% and 15% for CuO and graphite, respectively. The reason for that refer to the large difference in temperature between the basin plate and the ambient in the MSSs compared to CSS. These high losses have marginal impact on the SS thermal performance. Therefore, care must be taken during the design process of SS, especially in the MSSs, to reduce the thermal energy losses by using prober insulations with lower thermal conductivity.

The diurnal exergy efficiencies of the CSS components, i.e., the basin liner, the glazier cover and the saline water, are 11.89%, 17.66% and 63.61%, respectively; these values are increased for the CuO nanofluid still and reach 13.40%, 23.18% and 80.83% for the basin liner, the glazier and the saline water, respectively. Moreover, these values are increased for the graphite nanofluid still and reach 12.32%, 23.76% and 93.35% for the basin liner, the glazier cover and the saline water, respectively, as shown in Fig.5-b. Results shown in Fig.5-b are completely in agreement with thermodynamics laws that the higher exergy dissipation leads to lower energy and exergy efficiencies of the passive stills. This observation may be very useful for the further improvement in the design of stills to increase the cost-effective efficiencies leading to higher productivity.

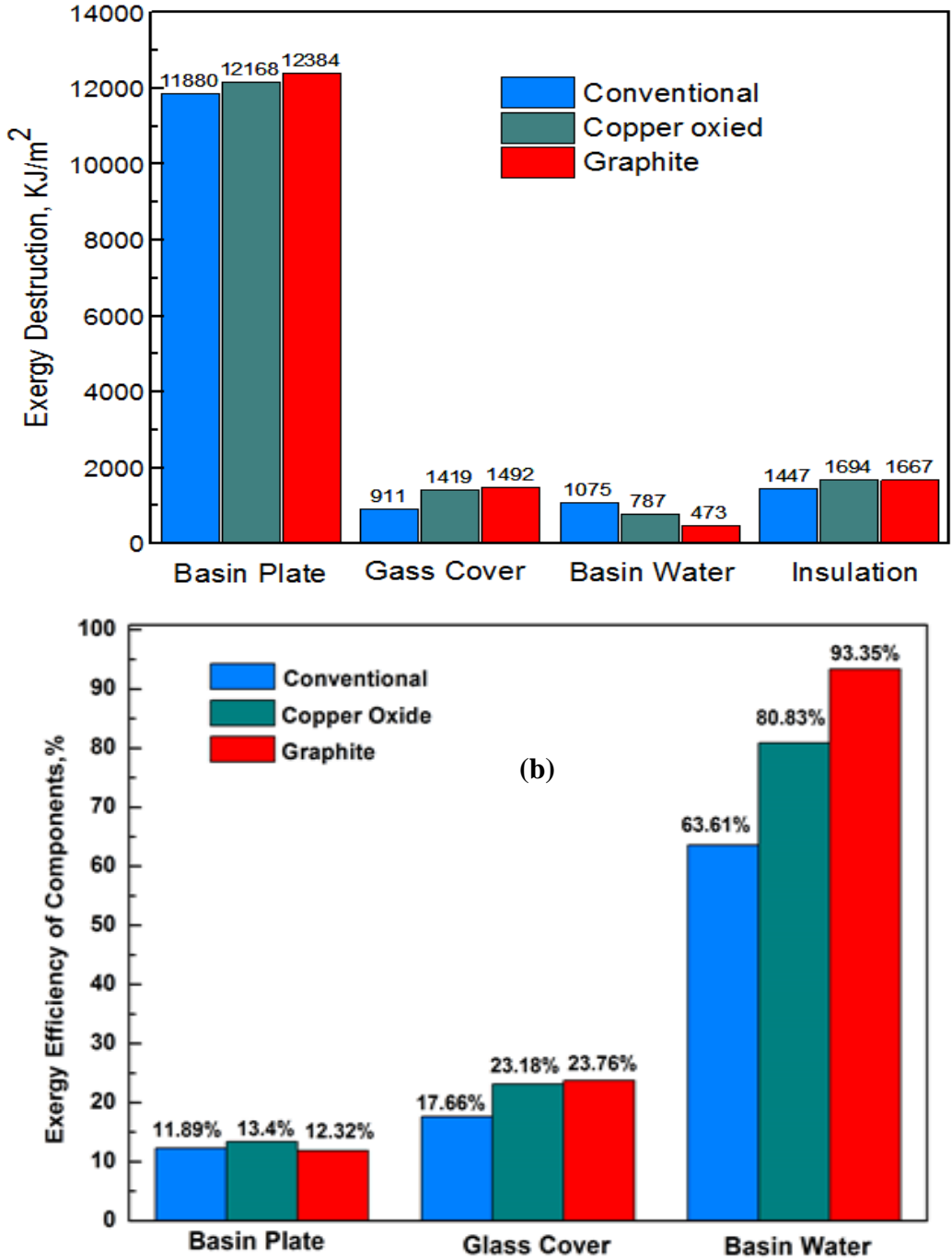

Fig. 5 (a) Daily variation of different parameters affects exergy dissipation. It shows that exergy destruction in the basin liner, the glass cover, the saline water and the insulation. (b) Diurnal exergy efficiencies of the still components.

## 4.6 Hourly, daytime energy and exergy efficiency

The hourly energy efficiency is illustrated in Fig. 6-a. The hourly energy efficiency increases gradually from the starting of the experiment up to the maximum value at approximately 14 pm for both still types understudy. The maximum hourly energy efficiency is 46.98% and 43.68% with graphite and CuO nanofluids, respectively, and 35.56% for the CSS (without nanofluid). Furthermore, the hourly exergy efficiency is illustrated in Fig. 6-b. The hourly exergy efficiency increases gradually from the starting of the experiment up to the maximum value at approximately 13 pm. This could be due to the stored heat energy inside the saline water during higher incident solar intensity period (9 am to 13 pm). The maximum hourly exergy efficiency is 6.06% and 5.79% with graphite and CuO nanofluids, respectively, and 4.32% for the CSS. Furthermore, it is observed that the energy efficiency is much higher than the exergy efficiency for both types of the SS understudy. This lower value is attributed to less available energy or low quality of evaporative thermal energy.

The daytime total exergy and energy efficiency for the MSSs and CSS are illustrated in Fig.6-c. The diurnal energy efficiencies according to equation (17) for modified stills (with graphite and CuO nanofluids) and the CSS are 41.18%, 38.61% and 29.17%, respectively. Furthermore, the diurnal exergy efficiencies according to equation (32) for modified stills (with graphite and CuO nanofluids) and the CSS are 4.32%, 3.78% and 2.63%, respectively. So, it is recommended to use nanofluids in the stills, as it enhances the total thermal performance.

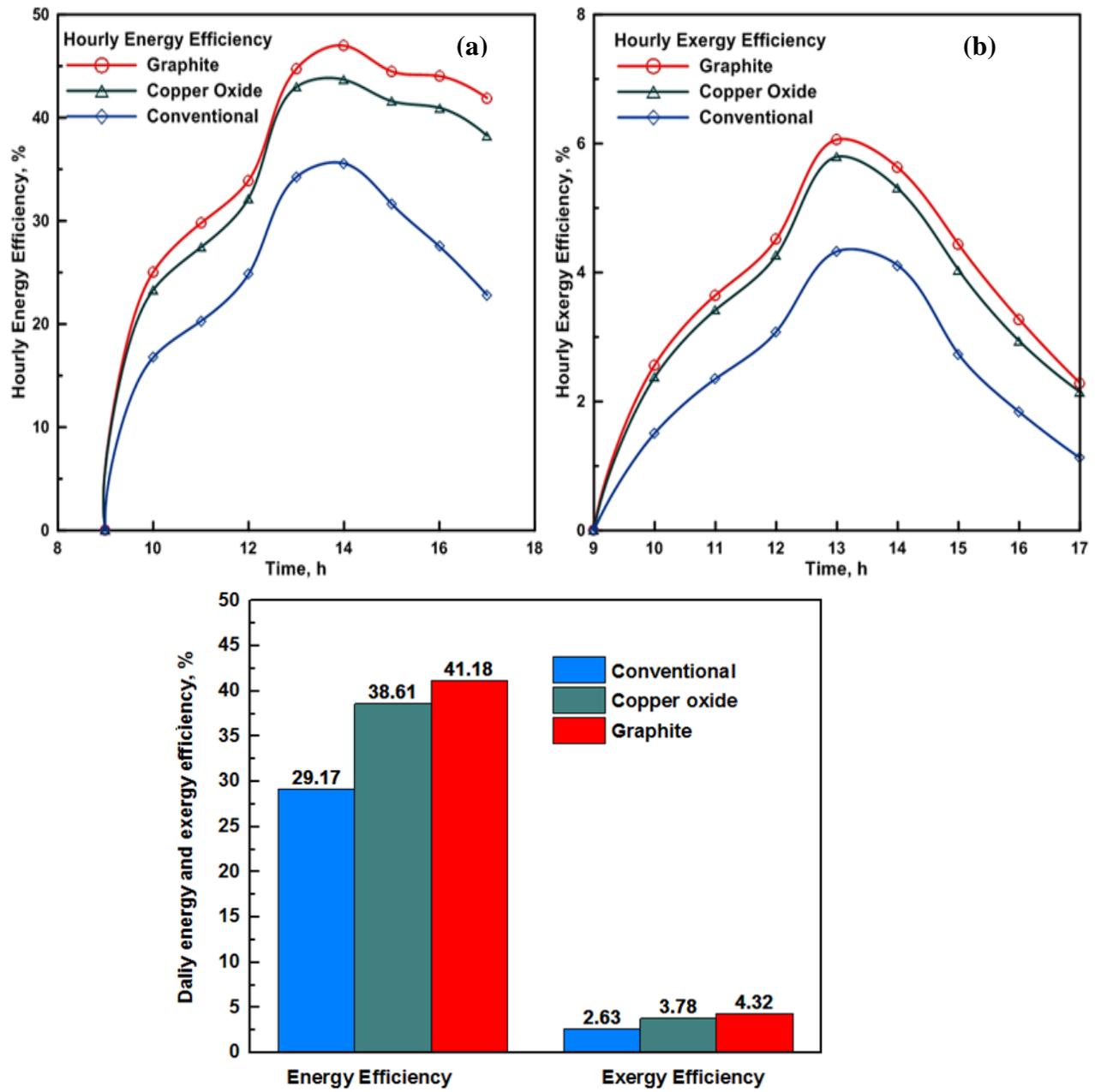

Fig.6. (a) Hourly energy efficiency, (b) Hourly exergy efficiency and (c) Daytime exergy and energy efficiency.

## 5. Conclusions

In this study, three solar stills are fabricated and analyzed to figure out the influence of utilizing nanoparticles (graphite and copper oxide) on the thermal performance of MSSs (with nanofluids) compared with a CSS (without nanofluids) at the same operating conditions. Firstly, the hourly and the diurnal and the total energy and exergy efficiencies are presented. The diurnal exergy efficiency is lower than the diurnal energy efficiency for all types of SSs understudy. Moreover, the energy and exergy efficiency are higher for MSSs in comparison with CSS. Secondly, the exergy destructions for all components of the stills understudy (basin, glazier and water) are analyzed. It is observed that among all SS components the largest exergy destruction occurs in the basin. The exergy destruction of basin has to be decreased by selecting suitable materials for the basin liner and the insulation which leads to enhance the exergy efficiency of solar still.

Furthermore, the values of results are concluded as:

- The diurnal exergy efficiency of graphite, CuO nanofluids and CSS are 4.32%, 3.78% and 2.63%, respectively.
- The diurnal exergy efficiencies of the CSS components, i.e., the basin, the glazier and the water, are 11.89%, 17.66% and 63.61%, respectively.
- The diurnal exergy efficiencies of the CuO nanofluid still components, i.e., the basin, the glazier and the saline water, are 13.40%, 23.18% and 80.83%.
- The diurnal exergy efficiencies of the graphite nanofluid still components, i.e., the basin liner, the glazier and the saline water, are. 32%, 23.76% and 93.35%

- The output of the stills with graphite and CuO nanofluid reached approximately 41.18% and 32.35 %, respectively, over the classical one.

- The diurnal energy efficiency of graphite and CuO nanofluids is 41.18% and 38.61%, respectively, and for CSS is 29.17%.

**Acknowledgement**

N.Y. was sponsored by National Natural Science Foundation of China (No. 51576076 and No. 51711540031), Hubei Provincial Natural Science Foundation of China (2017CFA046) and Fundamental Research Funds for the Central Universities (2016YXZD006). The authors thank the National Supercomputing Center in Tianjin (NSCC-TJ) and China Scientific Computing Grid (ScGrid) for providing assistance in computations.